\pdfoutput=1 

\documentclass[journal]{IEEEtran}

\usepackage{graphicx}
\usepackage[export]{adjustbox}
\usepackage[table,xcdraw]{xcolor}
\usepackage{multirow}
\usepackage{amsmath}
\usepackage{amssymb}
\usepackage{slashbox}
\usepackage{textcomp}
\usepackage{xspace}
\usepackage[T1]{fontenc}
\usepackage{comment}

\begin{document}
\title{PowerNet: Neural Electricity Demand Forecasting}


\author{Yao~Cheng, 
Chang~Xu, 
Daisuke~Mashima, 
Bin Zhou,
Vrizlynn~L.~L.~Thing, 
Yongdong~Wu
\thanks{Yao Cheng, Vrizlynn L. L. Thing, and Yongdong Wu are with the Institute for Infocomm Research, A*STAR, Singapore, email: \{cheng\_yao, vriz, wydong\}@i2r.a-star.edu.sg}
\thanks{Chang Xu was with the School of Computing Science and Engineering, Nanyang Technological University, Singapore, email: xuch0007@e.ntu.edu.sg.}
\thanks{Daisuke Mashima and Bin Zhou are with the Advanced Digital Sciences Center, Singapore. Email: \{daisuke.m, zhou.bin\}@adsc-create.edu.sg}}


\maketitle

\begin{abstract}
Power demand forecasting is a critical task for achieving efficiency and reliability in power grid operation. Accurate forecasting allows grid operators to better maintain the balance of supply and demand as well as to optimize operational cost for generation and transmission.
This article proposes a novel neural network architecture \textit{PowerNet}, which can incorporate multiple heterogeneous features, such as historical energy consumption data, weather data, and calendar information, for the power demand forecasting task. 
Compared to two recent works based on Gradient Boosting Tree (GBT) and Support Vector Regression (SVR), PowerNet demonstrates a decrease of 
33.3\% and 14.3\% 
in forecasting error, respectively.
We further provide empirical results the two operational considerations that are crucial when using PowerNet in practice, 
i.e., how far in the future the model can forecast with a decent accuracy and how often we should re-train the forecasting model to retain its modeling capability. 
Finally, we briefly discuss a multilayer anomaly detection approach based on PowerNet.
\end{abstract}


\section{Introduction}

The modern smart grid is an enhanced electrical grid that takes advantage of sensing and information communication technologies to improve the efficiency, reliability, and security of traditional power grid. 
Compared to the traditional power grid, entities in smart grids are able to obtain timely power grid status of many kinds.
Smart metering, which is a major improvement brought by smart grids, facilitates real-time metering and reporting of electricity consumption data. 
One resulting benefit is that the accurate, fine-grained \textit{power demand forecasting} can be carried out based on such meter measurement, which affects the power generation scheduling and power dispatching for a future period by predicting the power demand in that period using the historical data in hand. 

Demand forecasting is important in demand management for both power companies and electricity customers~\cite{siano2014demand}.
For power companies, based on the demand forecasting results, they can allocate proper resources to balance the supply and demand, or adjust the demand response strategy such as dynamic pricing to shape the load so as to avoid the infrastructure capacity strain or to avoid additional cost for starting peaker plants.
In addition, they can detect the abnormal meter measurements caused either by the unexpected meter failure or the deliberate meter manipulation by identifying those measurements that do not present a conformance to the predicted/expected values. 
For the electricity customers, power forecasting provides them with their expected power consumption and cost in a future period under dynamic pricing strategy, so that they can adjust their usage schedule accordingly to achieve a lower cost.

Although demand forecasting has been widely studied for years, a challenge in making accurate forecasting is that the power demand is subject to various influential factors which may have discriminative capability in influencing the power demand.
With this challenge in mind, we propose a novel forecasting neural network architecture named \textit{PowerNet}.
We take into account a set of features from three heterogeneous dimensions, i.e., the historical consumption data, the weather information, and the calendar information, all of which are considered influential on electricity customers' power consumption patterns. 
In each dimension, a set of features is developed. 
Then, we introduce our model, PowerNet, which is capable of incorporating all the designed features.
The key property of PowerNet is the ability to model both sequential data (i.e., historical consumption data) and non-sequential data (i.e., weather \& calendar information) in a unified manner.
The underpinning idea lies in the use of recurrent neural network for encoding dependencies implied in sequential data and multilayer perceptron network for capturing correlations between non-sequential features and predictions.
In order to evaluate the effectiveness of our model, we compare PowerNet with two state-of-the-art demand forecasting techniques based on Gradient Boosting Tree (GBT)~\cite{bansal2015energy} and Support Vector Regression (SVR)~\cite{yu2015towards}, respectively.
%
Moreover, we then tackle two crucial questions that need to be answered when operating PowerNet in practice:
how far in the future the model can forecast with a decent accuracy and how often we should re-train the forecasting model to retain its modeling capability.
Last but not the least, we discuss a multilayer data-driven anomaly detection approach based on PowerNet. 

The contributions of this work are summarized below.
\begin{itemize}
\item We propose PowerNet, a novel power demand forecasting neural network that captures heterogeneous features in a unified way.
\item We compare PowerNet with two representative models adopted in recent research works, i.e., GBT and SVR.
The results reveal that PowerNet reduces the Mean Square Error (MSE) by 33.3\% and 14.3\% compared to GBT and SVR, respectively.
\item We further evaluate the forecasting model under different forecasting duration and re-training frequency, using publicly available datasets. Our findings include: 
\begin{itemize}
\item PowerNet can serve the day-ahead forecasting tasks well. 
The Mean Absolute Percentage Error (MAPE) of the 24-hour forecasting grows over time but is capped at 10\%.
\item The effectiveness of our model after one training process can last 550 hours with MAPE around 11\% and 36 hours with MAPE less than 10\%. 
\end{itemize} 
\end{itemize}

The rest of this paper is organized as follows. In Section~\ref{sec:feature}, we discuss features to be incorporated into power demand forecasting. Section~\ref{sec:powernet} elaborates the design of PowerNet. We discuss evaluation results, including comparison with state-of-the-art techniques and empirical results to answer the aforementioned questions for practical operation in Section~\ref{sec:evaluation}, followed by a brief discussion about the application for anomaly detection in Section~\ref{sec:anomaly}. Related work is discussed in Section~\ref{sec: Related Work}, and we conclude the paper in Section~\ref{sec:concl}.

\section{Feature Design and Dataset}\label{sec:feature}

Power consumption patterns are affected by a variety of factors. Thus demand forecasting mechanism should incorporate such factors as features, in addition to historical energy consumption data. We focus on weather and calendar data. Below, we elaborate these features and dataset we utilize in this paper.

\begin{figure}
\centering
\includegraphics[scale=0.36]{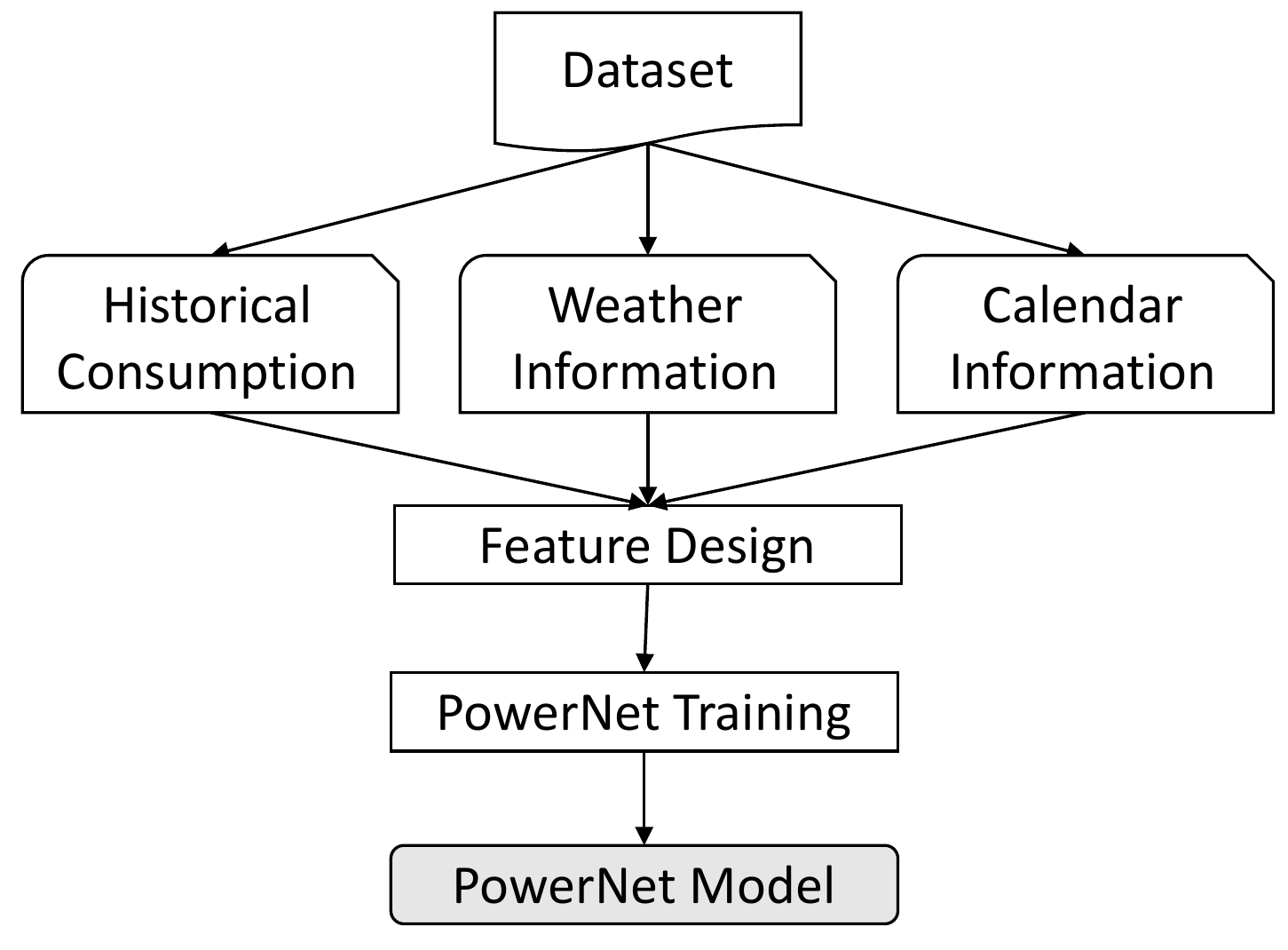}
\caption{Approach Overview.}
\label{fig: Approach Overview.}
\end{figure}

\subsection{Energy Usage Dataset}
\label{sec: Energy Usage Dataset}
We use the publicly available dataset provided by University of Massachusetts~\cite{dataset}.
It includes two parts, the apartment dataset and the weather dataset.

The apartment dataset contains data for 114 single-family apartments located in Western Massachusetts for the period from the year 2014 to 2016.
The dataset records the power of every single apartment in fixed temporal frequency\footnote{Given the metering interval is fixed, power values are able to represent the power consumption.}.
The metering frequency is once every 15 minutes for the year 2014 and 2015 (before December 15), and once every 1 minute for the year 2016. 
The data is in \textit{.csv} files, each of which records the power consumption details for one apartment within one year with apartment ID as its file name.
%
%

Along with the power consumption data, hourly weather information during the record period from 2014 to 2016 is available.
Fourteen meteorological attributes are included in the weather dataset including weather summary, temperature, humidity, cloud cover, wind speed, wind bearing, visibility, pressure, etc.
In our experiment, we use the data of the latest year 2016 because of its finer granularity in recording frequency as well as the latest consumption pattern it may reflect.


\subsection{Feature Design}

The features used by the existing forecasting models fall into three categories in terms of privacy issue, i.e., publicly available information (e.g., weather information and calendar information), household private information (e.g., demography), and quasi-private information (e.g., historical consumption data acquired by power utility companies).
The quasi-private information here is defined as privacy-related but not public available data.
For example, the historical consumption data can be used to infer certain private household characteristics~\cite{anderson2016electricity}, but it is only available to the authorized personnel within power utility companies instead of to the public.

Though it is natural for private household data to have a direct influence on the household power demand, e.g., more people living in the house leads to more power demand, in this work, we limit the predictors to non-private information due to the following reasons.
First of all, we would like to involve no household-specific data in forecasting procedure other than power meter readings due to user privacy concern.
Secondly, some utility companies may have access to household private data such as locations.
However, it is not common for utility companies to have other private information, for example, the demography information. 
Thirdly, the forecasting model not based on the house specific data can be applied to larger scales easily, such as building level or area level.

We develop three categories of features from the dataset, i.e., historical consumption data, weather information, and calendar information. 
Historical consumption data is the actual observation of the prediction target, which directly reflects the consumption pattern. 
Power utility companies can get this data by reading power meters. 
Weather information has an influence on the power demand since some appliances are sensitive towards weather conditions.
For example, the use of air conditioner depends on the temperature and humidity. 
Calendar information, such as weekday or weekend, shapes the user consumption behavior in terms of different living/working styles. 
It indicates the consumption pattern according to the calendar feature and cycle.

Our features based on the above three categories are summarized in Table~\ref{table: Prediction features.}.
There are $n+18$ features in total, among which, $n$ features are from historical consumption data, 13 are from weather information, and 5 are designed from calendar information.
The historical data involves a large number of data points.
Therefore, it is necessary to find out $n$ historical data points that are most correlated with the target forecasting value.
To solve this problem, we use AutoCorrelation Function (ACF), which can quantify the correlation between data points of various time lags, to find out the most related number of lag values $n$.

\begin{table*}[]
\centering
\caption{Features for the power demand forecasting task.}
\label{table: Prediction features.}
\scalebox{1}
{
\begin{tabular}{|c|l|}
\hline
\textbf{Category}  & \multicolumn{1}{c|}{\textbf{Detail}}                                                                                                                                                                         \\ \hline
Historical Consumption Data & Consumption data in past $n$ time slots \\ \hline
Weather Information         & \begin{tabular}[c]{@{}l@{}}Summary, icon, temperature, apparent temperature, cloud cover, precip probability, precip intensity, visibility,  \\ wind speed, wind bearing, humidity, pressure, dew point\end{tabular} \\ \hline
Calendar Information        & \begin{tabular}[c]{@{}l@{}}Day of the month, day of the week, hour of the day, period of the day (i.e., daytime and night time), \\ is weekend (boolean value)\end{tabular}                             \\ \hline
\end{tabular}
}
\end{table*}

%
%

\section{PowerNet}\label{sec:powernet}

\subsection{Overview}
\begin{figure*}[t]
	\centering
	\includegraphics[scale=0.43]{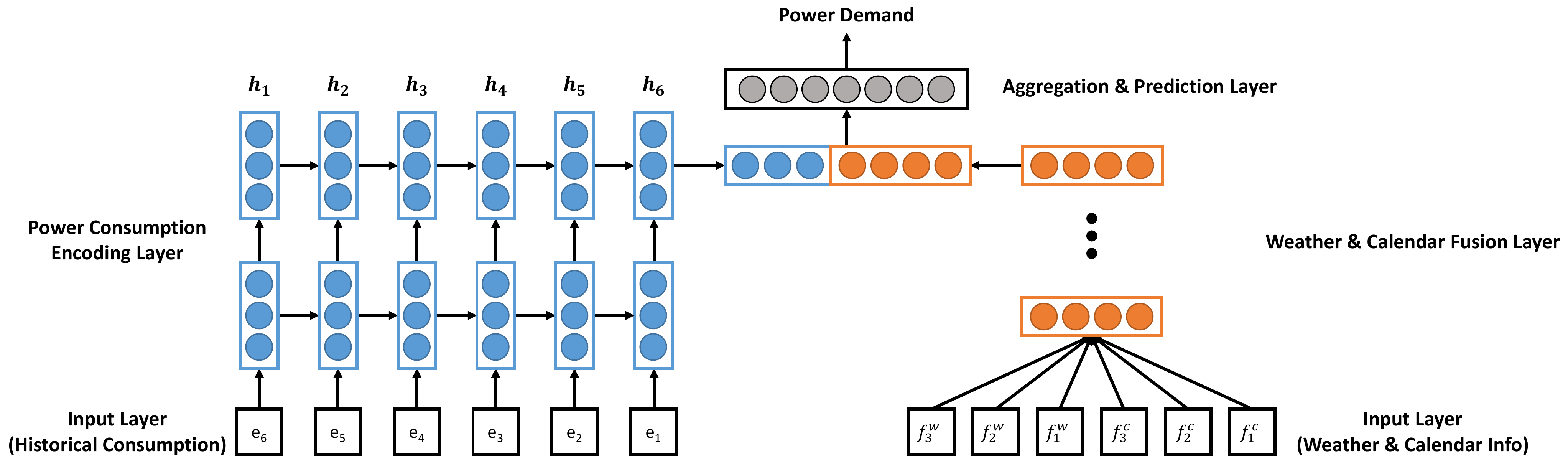}
	\caption{The Architecture of PowerNet.}
	\label{fig: architecture}
\end{figure*}

Our approach is to forecast power demand by modeling the relationship between power demand and relevant features. 
Fig.~\ref{fig: Approach Overview.} illustrates the high-level pipeline of our approach, including feature design discussed in Section~\ref{sec:feature}.
We propose a unified neural network model, named PowerNet, to jointly exploit the three categories of features developed in the previous section.
Figure~\ref{fig: architecture} shows the architecture of PowerNet.
It has two main components.
The left component (in blue) is designed to model the historical consumption time series data.
The key is to capture the temporal effects of power consumption in that future consumption could be correlated to consumption in the recent past.
Here, we utilize the Long Short-Term Memory Network (LSTM)~\cite{hochreiter1997long} to \textit{encode} the correlations between consecutive power consumption in time.
The right component (in orange) is a Multilayer Perceptron model (MLP)~\cite{hornik1989multilayer} that is capable of modeling the non-linearity in the weather and calendar data.
Finally, we aggregate the outputs from these two components and make ultimate predictions of power demand through a Prediction Layer.
In the following, we dissect each component of PowerNet.

\subsection{Input Layer}

To incorporate sequential data and non-sequential data, the input layer of PowerNet consists of two parts: one for the former and the other for the latter. 
The first part of input is a series of historical power consumption data $E=\{e_1,...,e_t,...,e_{|E|}\}$ where each entry $e_t\in\mathbb{R}^+$ is a real-valued non-negative power meter reading at time $t$.
The second part of input is the feature vectors of weather and calendar data, denoted by $\textbf{f}^w=\{f_1^w,...,f_{|w|}^w\}$ and $\textbf{f}^c=\{f_1^c,...,f_{|c|}^c\}$, where $|w|$ and $|c|$ equal to the numbers of weather and calendar features introduced.

\subsection{Power Consumption Encoding Layer}

The utility of this layer is to encode the power consumption time series data based on LSTM network, which is a variation of recurrent neural network that can learn long-term dependencies.
Different from traditional neural networks that can only take previous $N$ history readings as input, LSTM allows unlimited history information to persist with an internal loop mechanism while avoids the gradient vanishing problem~\cite{greff2016lstm}.
Therefore, it has been successfully applied to various areas, e.g., continual prediction~\cite{gers2000learning}, language modeling~\cite{mikolov2010recurrent}, and translation~\cite{sutskever2014sequence}.
The core of LSTM is a memory cell that can maintain information across time via gating mechanism.
The LSTM cell maintains a cell status based on both current input $x_t$ and previous output $h_{t-1}$ (i.e., the recurrent input), and then decides what information to be left and what to be passed on (i.e., $h_t$).
We do not detail the gating mechanisms here which can be found in previous literature~\cite{hochreiter1997long}.
We use \textit{LSTM()} to represent the cell function.

In PowerNet, we apply a stacked LSTM to every time step of the power consumption time series data $E$,
\begin{equation}
\begin{split}
[h_1\,\,c_1]&=\text{LSTM}^{\text{stack}}(e_1,h_0,c_0)\\
&...\\
[h_t\,\,c_t]&=\text{LSTM}^{\text{stack}}(e_t,h_{t-1},c_{t-1})\\
&...\\
[h_{|E|}\,\,c_{|E|}]&=\text{LSTM}^{\text{stack}}(e_{|E|},h_{|E|-1},c_{|E|-1})
\end{split}
\end{equation}
Finally, the output of $\text{LSTM}^{\text{stack}}$ at the last time step $h_{|E|}\in\mathbb{R}^{n}$ is used as a ultimate encoding of the entire power consumption series, where $n$ is the LSTM memory size.


\subsection{Weather \& Calendar Fusion Layer}

In this layer, we handle input from the weather \& calendar features. 
Specifically, we jointly model the two feature vectors through a multilayer perceptron network (MLP),
\begin{equation}
o=\text{ReLU}(W_2\text{ReLU}(W_1[\textbf{f}^w;\textbf{f}^c]+b_1)+b_2)
\end{equation}
where $W_1\in\mathbb{R}^{d_1\times m},W_2\in\mathbb{R}^{d_2\times d_1},b_1\in\mathbb{R}^{d_1},b_2\in\mathbb{R}^{d_2}$ are trainable weights, $m=|w|+|c|$, $d_1,d_2$ are the sizes of hidden units, $[;]$ denotes vector concatenation by column, and $o\in\mathbb{R}^{d_2}$ is the output encoding of this MLP.
ReLU~\cite{nair2010rectified} is used as the activation function for introducing non-linearity.

\subsection{Aggregation \& Prediction Layer}

Having both power consumption history and weather \& calendar information encoded, we aggregate the obtained encodings and make the final predictions.
Concretely, we concatenate the two encodings $h_{|E|}$ and $o$ and feed the result through a final feed-forward regression network,
\begin{equation}
\hat{y}=W_4\text{ReLU}(W_3[h_{|E|};o]+b_3)+b_4
\end{equation}
where $W_3\in\mathbb{R}^{d_3\times (|E|+d_2)},b_3\in\mathbb{R}^{d_3},W_4\in\mathbb{R}^{1\times d_3},b_4\in\mathbb{R}$ are trainable parameters and $d_3$ is the hidden size of the inner layer.
Note that both $W_4$ and $b_4$ of the outer layer have only one hidden unit for producing the final predicted reading value.
$\hat{y}\in\mathbb{R}$ is the predicted power consumption reading value.

\subsection{Optimization}
For model training, we use mean squared error loss (Eq.~(\ref{eq:loss})) with dropout regularization~\cite{srivastava2014dropout},
\begin{equation}
\label{eq:loss}
L(W_{\ast},b_{\ast})=\frac{1}{N}\sum_{i=1}^{N}(\hat{y}_i-y_i)^2
\end{equation}
where $N$ is the number of training examples, $W_{\ast},b_{\ast}$ are all the aforementioned trainable parameters in our model.
In addition, all trainable parameters in the fully-connected layers are regularized by L2 norm.
Finally, adam (Adaptive Moment Estimation)~\cite{kingma2014adam} is used as the optimizer for stochastic gradient descent.

\section{Evaluation}\label{sec:evaluation}

This section first compares PowerNet with two representative models used in recent works~\cite{bansal2015energy}\cite{yu2015towards} in terms of two quantitative metrics.
Then, we evaluate PowerNet under different settings, including the forecasting frequencies, forecasting periods, and the freshness of PowerNet.

\subsection{Preparation}
\subsubsection{Baseline}

We select two recent works as our baseline models in this work.
Technically, one of them adopts GBT~\cite{bansal2015energy} and the other one adopts SVR~\cite{yu2015towards}.
For a fair comparison, we implement their models as well and apply the implemented models to the same public dataset as described in Section~\ref{sec: Energy Usage Dataset}.

GBT is adopted by Bansal et al.~\cite{bansal2015energy} to forecast power consumption.
GBT is a supervised learning predictive model which can be used for classification and regression purpose~\cite{friedman2001greedy}\cite{friedman2002stochastic}.
GBT builds the model, i.e., a series of trees, in a step-wise manner.
In each step, it adds one tree, and maintains the existing trees unchanged.
The added tree is the optimal tree by minimizing a predefined loss function. 
Basically, GBT is an ensemble of weaker prediction models, which becomes a better model, which is exactly the core idea of the gradient.

SVM is used in the work by Yu et al.~\cite{yu2015towards} to forecast power usage.
SVM is a supervised machine learning algorithm for solving both classification and regression problems~\cite{boser1992training}.
SVM does classification by seeking the hyperplane that differentiates the two classes to the largest extent, i.e., maximizing the margin. 
Similarly, regression using SVM is called SVR~\cite{muller1997predicting} is to seek and optimize the generation bounds by minimizing the predefined error function. 
The regression can be conducted in both linear and non-linear manner. 
For the non-linear SVR, it needs to transform the data into a higher dimensional space so that it is possible to perform the linear separation.

\subsubsection{Evaluation Metric}

We introduce two metrics to evaluate the accuracy of the forecasting model, i.e., Mean Square Error (MSE) and Mean Absolute Percentage Error (MAPE). 
The smaller the error is, the more accurately the model predicts.

MSE measures the average of the square errors/deviations as directed by Equation~\ref{eq:MSE}.
$n$ is the total number of forecasting values, $A_t$ denotes the actual value at time $t$, and $F_t$ denotes the forecasting value at time $t$. 
The closer the value to zero, the better the prediction is.

\begin{equation}\label{eq:MSE}
MSE=\frac{1}{n}\sum_{t=1}^{n} (A_{t}-F_{t})^2
\end{equation}

Different from MSE, MAPE measures the error proportion to the absolute value.
It expresses the error as a percentage and can be calculated using Equation~\ref{eq:MAPE}.

\begin{equation}\label{eq:MAPE}
MAPE = \frac{100\%}{n}\sum_{t=1}^{n} \left | \frac{A_{t}-F_{t}}{A_{t}} \right |
\end{equation}

MSE is more useful in comparison experiments with identical test data, as it is the absolute square error value which depends on the scale of actual values.
Comparing to MSE, MAPE is more indicative in the comparison between different data since it represents the error in a percentage manner.
\subsection{Comparison with Baselines}

In this experiment, we compare our model with two recent works, i.e., the works of Bansal et al.~\cite{bansal2015energy} and Yu et al.~\cite{yu2015towards} under the identical setting, with the same training and testing data.
The two works~\cite{bansal2015energy}~\cite{yu2015towards} are referred to as ``GBT'' and ``SVR'' for short in this section, respectively.
Our PowerNet uses a two-layered LSTM network. 
The cell memory size for every layer is tuned from the set \{64, 128, 256, 512\} using grid search.
Early stopping is employed when there is no further improvement on the validation set.
Similarly, the parameters for baseline models are also automatically tuned in the same way. 
For GBT, three parameters are involved, i.e., the number of boosting stages to perform $n\_estimators$, maximum depth of the individual regression estimators $max\_depth$, and learning rate $learning\_rate$.
Its parameter grid is constructed using $n\_estimators$: \{50, 100, 150, 200, 250, 300, 350, 400, 450, 500\}, $max\_depth$: \{1, 2, 3, 4, 5\}, and $learning\_rate$: \{0.001, 0.01, 0.1, 1\}.
For SVR, three parameters $C$, $kernel$ and $gamma$ are involved.
We construct the parameter grid using $C$: \{0.001, 0.01, 0.1, 1\},  $kernel$: \{rbf, linear, poly, sigmoid\}, and hence $gamma$ is automatically set to the corresponding kernel coefficient or the reciprocal of the number of features.

We use the power consumption data of past 26 days, i.e., 624 hours as the training set to train the three models, and the next 48 hours data, i.e., day 27-28 as the validation set.
Finally, we make predictions on the test data of day 29-30.
Due to the space limit, we only demonstrate the results of our model and the two baselines on the data of a randomly chosen apartment (No. 69 in April).
In particular, the results are obtained by training on the data from 1st April to 26th April (validating on data of 27th and 28th April) and testing on the data of 29th and 30th April.
As Fig.~\ref{fig: Forecasting accuracy of DPDF, SVR and GBT} shown, our model is able to capture the trend as well as peaks and valleys better than both two other works do.
PowerNet brings a decrease in both MSE and MAPE as shown in Table~\ref{table: Effectiveness of forecasting}.
It decreases 33.3\% and 14.3\% in MSE compared to GBT~\cite{bansal2015energy} and SVR~\cite{yu2015towards}, respectively.

\begin{figure}
  \centering
  \includegraphics[scale=0.49]{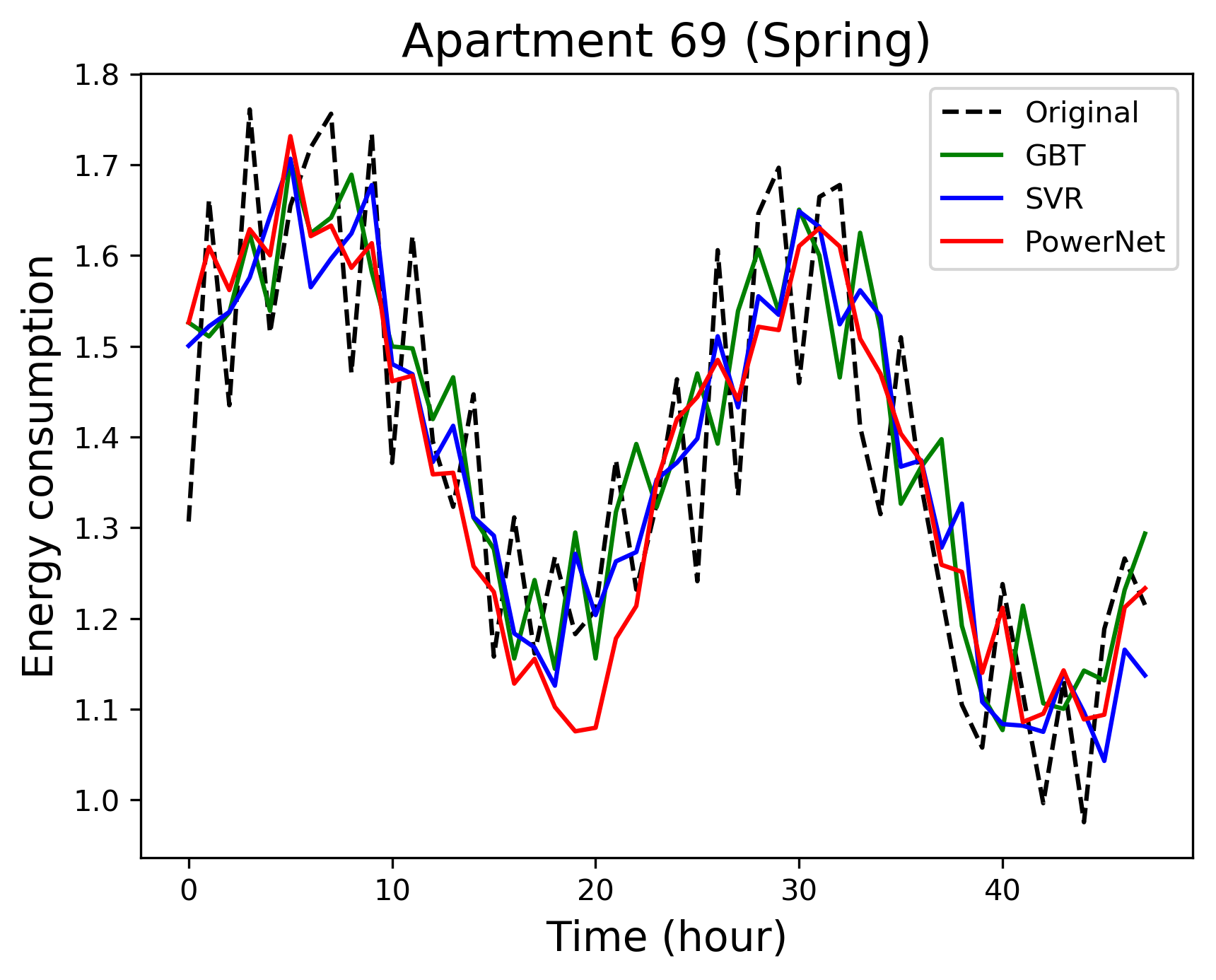}
  \caption{Model comparison results from GBT~\cite{bansal2015energy}, SVR~\cite{yu2015towards} and PowerNet.}
  \label{fig: Forecasting accuracy of DPDF, SVR and GBT}
\end{figure}

\begin{table}[]
\centering
\caption{Accuracy of PowerNet, SVR and GBT.}
\label{table: Effectiveness of forecasting}
\scalebox{1}{
\begin{tabular}{|c|c|c|c|}
\hline
\backslashbox{Error}{Model}              & \textbf{PowerNet} & \textbf{SVR~\cite{yu2015towards}} & \textbf{GBT~\cite{bansal2015energy}} \\ \hline
\textbf{MSE}  & 0.012     & 0.014    & 0.018    \\ \hline
\textbf{MAPE} & 7.115\%       & 7.835\%     & 8.783\%      \\ \hline
\end{tabular}
}
\end{table}

\subsection{Forecasting Period of PowerNet}
\label{sec: Forecasting Period in the Future}
In general, the accuracy of power demand forecasting deteriorates as a forecasting period becomes longer. thus, it is crucial for grid operators to know how much time ahead PowerNet can predict the demand without facing significant accuracy drop. In this section, we provide empirical results on forecasting accuracy against different forecasting periods using the real-world electricity consumption data.  
%
By doing so, grid operators can evaluate whether PowerNet is applicable for certain tasks that require different lengths of prediction period, such as the bidding in the day-ahead electricity market and day-ahead electricity scheduling which require the forecasting results one day ahead~\cite{conejo2005day}.

Some features for predicting the power demand in the far future may not available at the time of prediction.
For example, the power consumption of the previous one hour is an important feature to predict the power demand for the next hour.
If we predict more than one hour at once, we cannot know the actual consumption value for every ``previous'' hour, since it is not known yet.
Therefore, the prediction in the far future relies on the predicted values previous to that.
It means that there is a risk of error accumulation.

\begin{figure*}
\centering
\includegraphics[scale=0.66]{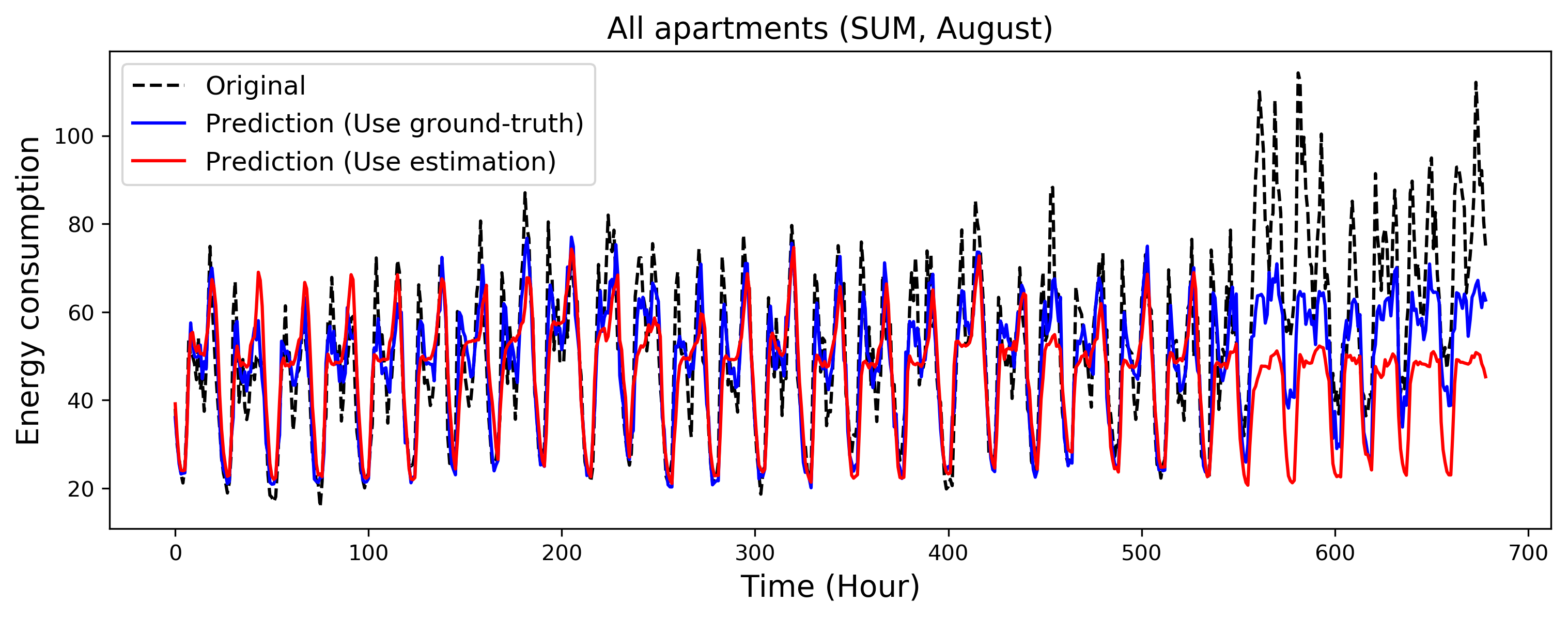}
\caption{Forecasting results using predicted and actual values.}
\label{fig: Prediction results.}
\end{figure*}

\begin{figure}
\centering
\includegraphics[scale=0.45]{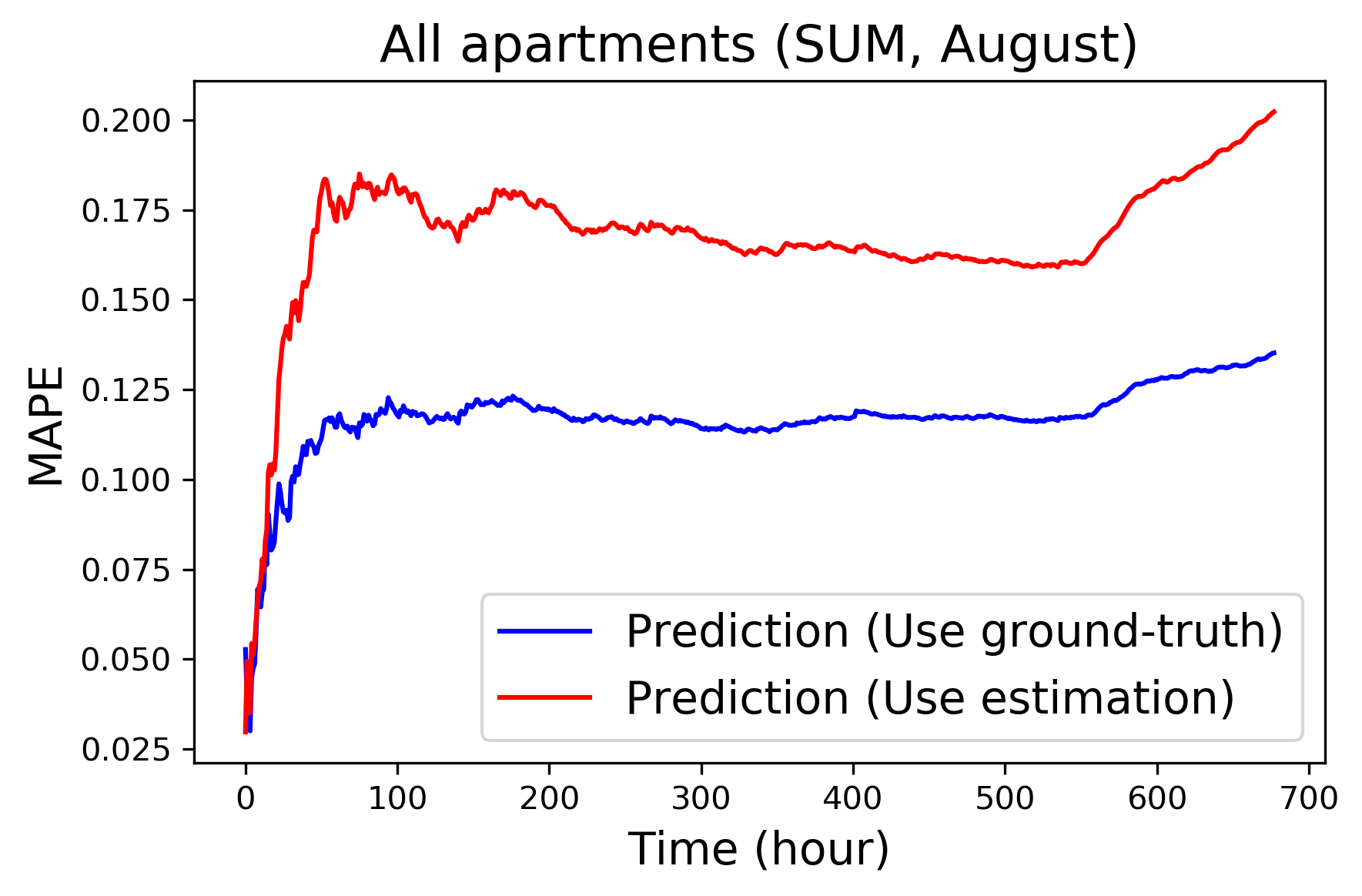}
\caption{Forecasting MAPE using predicted and actual values.}
\label{fig: Prediction MAPE.}
\end{figure}

In this experiment, we predict the power demand for the future 30 days at once based on current historical data.
We train the model on the aggregated historical data in June and predict the power demand for the following 30 days.
The forecasting results are shown in Fig.~\ref{fig: Prediction results.} in red.
We can see that the red line follows the original peaks and valleys well at the beginning. 
However, starting from some point around 550 on the x-axis, the red line totally loses track of the original values.
In order to understand the error quantitatively, we plot MAPE in Fig.~\ref{fig: Prediction MAPE.} in red.
We can see from the MAPE plot that the error increases as it goes further into the future.
Specifically, before 24 on the x-axis, the MAPE is at a low level less than 10\%. 
Then, MAPE rises a regional peak 18\% at 52 on the x-axis.
After that, MAPE declines a bit to 16\% and maintains the value till 550 on the x-axis from which the error increases sharply.
Given the experimental results, the model is suitable for forecasting in the day-ahead bidding task and day-ahead electricity scheduling. 


\subsection{Model Retraining Interval}
\label{sec: Model Validation Time}

For any data-driven model, it is necessary to keep the model up to date by retraining the model using fresh data. In particular, power consumption patterns are not stationary, and the trained model would become obsolete over time, which would result in lower forecasting accuracy. Thus, the timing for retraining is a crucial tuning parameter in real-world operation.
Retraining usually happens when degrading in prediction is noticed.
This subsection is to empirically investigate appropriate model retraining interval
how long a trained PowerNet model can be used with acceptable accuracy. 
It also provides us with insight on how often PowerNet should be re-trained to capture the new power demand characteristics evolved with time.

This experiment is different from the previous experiment in Section~\ref{sec: Forecasting Period in the Future}. 
The experiment in Section~\ref{sec: Forecasting Period in the Future} focuses on exploring the accuracy fluctuation caused by different lengths of forecasting periods, and it forecasts the power demand for a period at once based on the data on hand at that moment.
Differently, this experiment uses actual data, which eliminates the error accumulation caused by forecasting using estimated feature values.
We use the model trained in Section~\ref{sec: Forecasting Period in the Future}, and test it using the actual data in July.

The results are shown in Fig.~\ref{fig: Prediction results.} using the blue line.
Generally, the prediction based on actual values (the blue line) is better than the prediction based on predicted values (the red line), which is reasonable and as expected.
From the MAPE plot which is the blue line in Fig.~\ref{fig: Prediction MAPE.}, the same conclusion can be drawn. 
We can see the error increases at the beginning which aligns with the red line before 15 on the x-axis, and it keeps increasing to 10\% at 36 on the x-axis. 
Then, the error maintains around 11\% till 550.
At the very end, it reaches the largest error 13\%.
In practice, depending on the error tolerance of the prediction task, we can adjust our model by re-training the model with new data. 
For example, we can re-train the model every 36 hours to capture the new characteristics of the data generated during the 36 hours.
Generally, the model can maintain an MAPE around 11\% for more than 3 weeks.


\section{PowerNet for Anomaly Detection}\label{sec:anomaly}

Anomaly detection is to identify patterns in data that do not conform to the defined normal behavior~\cite{chandola2009anomaly}.
Anomaly detection in smart grids focuses on the non-technical loss which is not caused by the intrinsic loss (technical loss, e.g., transmission loss) in a power system.
Electricity theft is one of the most focused non-technical loss that causes anomalies.
Data-driven anomaly detection can be done by modeling the normal consumption behavior and defining a normal region.
Any consumption does not fall within the normal region is considered as an anomaly and potentially indicating a problem in the smart grid.
The forecasting results from PowerNet can be interpreted differently depending on the tasks, e.g., the power demand at some time in the future or the expected normal consumption at that time.
In the latter sense, PowerNet can be used to define the normal consumption behavior based on which further anomaly detection can be carried out.

\begin{figure*}
\begin{tabular}{cc}
\begin{minipage}[c]{0.48\hsize}
\centering
\includegraphics[scale=0.58]{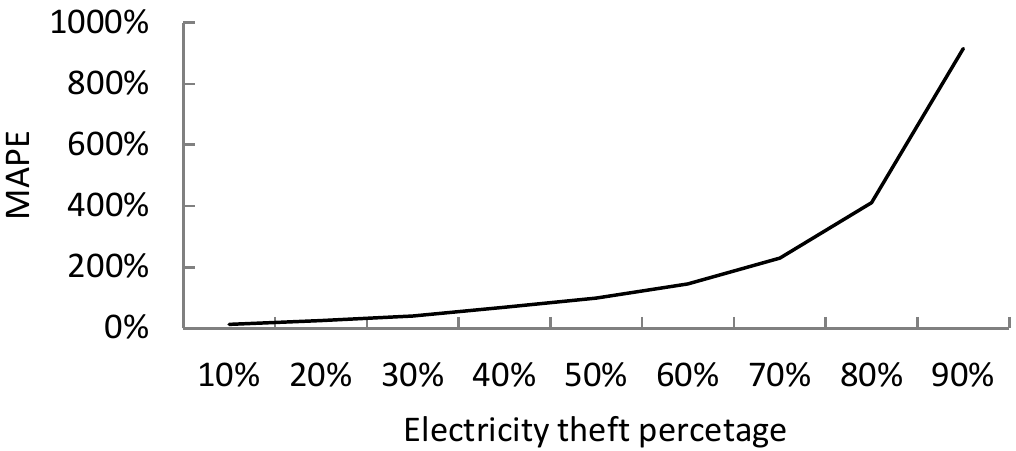} 
\caption{The MAPE predictions over different electricity theft scenarios characterized by the theft percentage from 10\% to 90\%.}
\label{fig:electricity_theft}
\end{minipage}
\begin{minipage}[c]{0.48\hsize}
\centering
\includegraphics[scale=0.58]{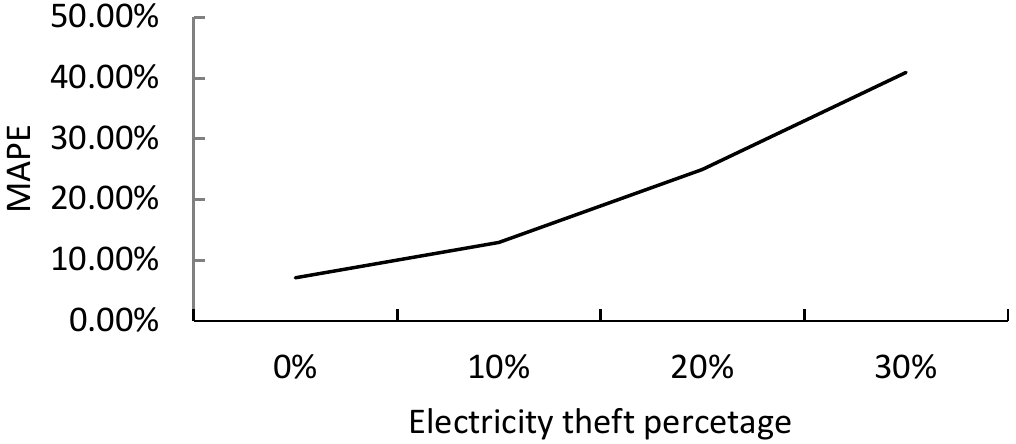} 
\caption{The MAPE predictions over different electricity theft scenarios characterized by the theft percentage from 10\% to 30\%.}
\label{fig:electricity_theft_mag}
\end{minipage}
\end{tabular}
\end{figure*}

Normally, for a consumer $u$, the reported consumption $M_r$ should be equal to the actual consumption $M_{u}$.
However, an attacker may be able to manipulate $M_r$ aiming at reducing the bill by making $M_r < M_u$.
We conduct a preliminary experiment to understand the performance of PowerNet when electricity theft happens.
We artificially reduce the power consumption by different theft percentages in the test data to simulate different electricity theft scenarios. 
Fig.~\ref{fig:electricity_theft} shows the forecasting MAPE results under different theft percentages and Fig.~\ref{fig:electricity_theft_mag} magnifies the the first 30\% of the x-axis in Fig.~\ref{fig:electricity_theft}. 
We can see from the magnified view (Fig.~\ref{fig:electricity_theft_mag}) that when theft percentage is small, the MAPE grows linearly as the percentage of theft grows.
However, from the experimental results in Fig.~\ref{fig:electricity_theft}, we can see that the overall MAPE increases in an exponential manner.
It means that the more the user steals, the larger the deviation between the predicted value $M_p$ and the reported value $M_r$ is.
In addition, the more the user steals, the more obvious the deviation is. 
A reasonable threshold that would trigger an alarm can be inferred from the historical statistic data as well as the tolerance of theft.

Anomaly detection can be deployed in both substation layer and individual consumer layer.
We discuss how PowerNet can be utilized to detect such anomalies in both layers.

\textbf{Anomaly detection in substation layer.}
On the substation level, there is a master meter which is a meter to measure the overall consumption of the whole supply region. 
The reading of master meter is denoted as $M_{s}$.
So we have $M_{s}=\sum_{i=1}^{n}M_{u}^i+TL$, where $n$ is the number of consumers in the supply region and $TL$ is the technical loss.
The substation can observe $M_{r}^i$ which is the reported consumption of consumer $i$. 
We can obtain $TL$ through $TL = M_{s}-\sum_{i=1}^{n}M_{u}^i$.
In normal case where $M_{r}^i = M_{u}^i$, we have $TL = M_{s}-\sum_{i=1}^{n}M_{r}^i$.
In order to detect the anomaly where $M_{r}^i \neq M_{u}^i$, we use PowerNet to model the indirectly observed $TL_o$.
In the attack case where $M_{r}^i \neq M_{u}^i$, a deviation would be observed between the predicted $TL_p$ and the observed $TL_o$.
Hence, PowerNet is able to detect the anomaly under a substation supply region by constructing one model for one substation.

\textbf{Anomaly detection in individual consumer layer.}
Anomaly detection on substation level can detect the anomaly but cannot determine which consumer is suspicious.
On the individual consumer level, with the help of the PowerNet, we can build a model for the consumer $u$ based on her historical $M_u$.
Once the attacker reduces her $M_r$ to make $M_r \neq M_u$, we shall notice that there is a deviation between her $M_r$ and $M_p$ which is predicted by PowerNet.
In this sense, anomaly detection in individual consumer layer can work as a complementary to anomaly detection in substation layer, which is able to locate the consumer who is suspiciously reporting false readings.

\section{Related Work}
\label{sec: Related Work}

The existing works on power demand forecasting can be generally classified into two categories, i.e., classic statistical models and modern machine learning algorithms.
 
In terms of statistical models, time-series models have been used to capture the time-series characteristics of power demand, e.g., ARMA~\cite{gross1987short}\cite{mashima2012evaluating}, ARIMA~\cite{alberg2017short}\cite{cho1995customer}.
Beside time-series models, Hong et al.~\cite{hong2010modeling} adopt multiple linear regression to model hourly energy demand using seasonality (regarding year, week, and day) and temperature information.
Their results indicate that complex featuring of the same information results in a more accurate forecasting. 
Fan and Hyndman~\cite{fan2012short} use the semi-parametric additive model to explore the non-linear relationship between energy usage data and variables, i.e., calendar variables, consumption observations, and temperatures, in the short-term time period. 
Their model demonstrates sensitivity towards the temperature.
In addition, conditional kernel density estimation is applied to the power demand forecasting area which performs well on the data with strong seasonality~\cite{arora2016forecasting}.
However, these models are limited in incorporating heterogeneous features in a unified way. 
Differently, the design of PowerNet makes it such a neural network that it is able to encode sequential features and single-value features simultaneously.

Regarding the machine-learning models, there are three models widely used for demand forecasting tasks, namely Decision Tree (DT)~\cite{bansal2015energy}\cite{gladysz2008application}\cite{yu2010decision}, Support Vector Machine (SVM)~\cite{yu2015towards}\cite{hong2011electric}\cite{qiu2013electricity}\cite{son2015forecasting},
and Artificial Neural Network (ANN)~\cite{gajowniczek2014short}\cite{zufferey2016forecasting}\cite{marino2016building}.
DT is used to predict building energy demand levels~\cite{yu2010decision} and analyze the electricity load level based on hourly observations of the electricity load and weather~\cite{gladysz2008application}.
Later, Bansal et al.~\cite{bansal2015energy} use the boosted DT to model and forecast energy consumption so as to create personalized electricity plans for residential consumers based on usage history.
There are also works using SVR, the regression based on SVM, to forecast power consumption in combination with other techniques, such as fuzzy-rough feature selection~\cite{son2015forecasting}, particle swarm optimization algorithms~\cite{qiu2013electricity}, and chaotic artificial bee colony algorithm~\cite{hong2011electric}.
The SVR-based prediction has demonstrated good prediction results~\cite{yu2015towards}.
For the third model ANN, Gajowniczek and Zabkowski choose ANN because they believe that time-series analysis is not suitable for their work since they observe high volatility in the data~\cite{gajowniczek2014short}. 
Zufferey et al.~\cite{zufferey2016forecasting} apply time delay neural network and find out that the individual consumer's consumption is harder to predict than an aggregation of multiple consumers. 
Recently, researchers take advantage of LSTM to forecast building energy load using historical consumption data~\cite{marino2016building}.
Cheng et al.~\cite{cheng2017powerlstm} further manage to feed the concatenation of historical data and influence features as a sequential input to the LSTM network.
Since they only use the LSTM network, all data are treated as sequential data.
Despite the extensive research carried out in power demand forecasting area, to the best of our knowledge, there is no such neural network architecture taking consideration of heterogeneous features as PowerNet does.

Another steam of related work is the anomaly detection in smart grids for non-technical loss such as electricity theft.
Bandim et.al~\cite{bandim2003identification} introduce an observer meter to observe the meter consumption of a set of users, and further identify the tampered meter using the deterministic and statistic approach.
Later, Krishna et al.~\cite{krishna2016f} discuss the detection capability based on such extra meters on different attacks.
Other than these, linear regression~\cite{liu2016regression}, cluster outlier~\cite{menon2016anomaly}\cite{chen2011energy} and SVM~\cite{nagi2010nontechnical}\cite{jokar2016electricity} are also used to detect the anomaly in smart girds.
Furthermore, Mashima et al.~\cite{mashima2012evaluating} evaluate the effectiveness of several anomaly detection models including the average detector, ARMA-GLR, and nonparametric statistics, and Local Outlier Factor (LOF).
In this work, we discuss that PowerNet can be used in multiple anomaly detection layers.



\section{Conclusions}\label{sec:concl}
In this article, we propose PowerNet, a power demand forecasting model based on modern recurrent neural network and multilayer perceptron network, which are capable of incorporating heterogeneous influence factors in a unified way.
It demonstrates improvement in prediction accuracy compared to two state-of-the-art approaches.
Further evaluation under different settings with the real-world dataset is carried out to better understand the model capability and crucial operational considerations in practice, namely the length of the forecasting period and the model retraining interval. 
Finally, we briefly discussed the potential of PowerNet being adopted in the anomaly detection task in the smart metering process.

%

\section*{Acknowledgement}
This research is supported by the National Research Foundation, Prime Minister's Office, Singapore under the Energy Programme and administrated by the Energy Market Authority (EP Award No. NRF2014EWT-EIRP002-040 and NRF2017EWT-EP003-047).

\bibliographystyle{IEEEtran}
\bibliography{prediction}

\end{document}